\documentclass[12pt]{article}

\usepackage[latin1]{inputenc}
\usepackage{amsmath}
\usepackage{amsfonts}
\usepackage{amssymb,slashed}
\usepackage{amsxtra}
\usepackage[margin=3cm]{geometry}
\usepackage{color}
\usepackage{hyperref}
\hypersetup{linktocpage=true}
\usepackage{graphicx}
\usepackage{placeins}
\usepackage{upgreek} 
\usepackage{mathrsfs}
\usepackage{accents}
\usepackage{multirow} 
\usepackage{tikz}

\numberwithin{equation}{section}
\newcommand{\beq}{\begin{equation}}
\newcommand{\eeq}{\end{equation}}
\def\be {\begin{equation}}
\def\ee {\end{equation}}
\def\bs#1\es{\begin{split}#1\end{split}}
\def\ba#1\ea{\begin{align}#1\end{align}}
\def\baed#1\eaed{\begin{aligned}#1\end{aligned}}
\def\bged#1\eged{\begin{gathered}#1\end{gathered}}
\def\bea{\begin{eqnarray}}
\def\eea{\end{eqnarray}}

\def\nn{\nonumber}

\def\d{\delta}
\def\D{\Delta}
\def\e{\epsilon}

\def\f{\phi}

\def\F{\Phi}
\def\g{\gamma}
\def\G{\Gamma}
\def\h{\eta}
\def\k{\kappa}

\def\m{\mu}
\def\n{\nu}

\def\O{\Omega}
\def\p{\psi}

\def\r{\rho}
\def\s{\sigma}

\newcommand{\cC}{\mathcal{C}}

\newcommand{\cW}{\mathcal{W}}

\newcommand{\cR}{\mathcal{R}}

\def\cO{{{\mathcal O}}}
\def\cM{\mathcal{M}} 
\def\cN{\mathcal{N}}

\def\bR{\mathbb{R}}

\def\Tr{\text{Tr}}

\def\pa{\partial}
\def\na{\nabla}
\def\fr{\frac}
\def\tfr{\tfrac}
\def\id{\rlap 1\mkern4mu{\rm l}}
\def\we{\wedge}

\def\lra{\leftrightarrow}

\def\tbzero{{\text{\tiny{(0)}}}}
\def\tbone{{\text{\tiny{(1)}}}}
\def\tbtwo{{\text{\tiny{(2)}}}}
\def\eppr{\alpha}

\newcommand{\wh}[1]{ {\hat{#1}}{} }
\newcommand{\til}[1]{ {\tilde{#1}} }

\let\foo\bar 
\renewcommand{\bar}[1]{ {\foo{  #1} }{} }

\newlength{\dhatheight}

\begin{document}

\baselineskip=16pt
\setlength{\parskip}{6pt}

\begin{titlepage}
\begin{flushright}
\parbox[t]{1.4in}{
\flushright MPP-2014-329}
\end{flushright}

\begin{center}

\vspace*{1.5cm}

{\Large \bf  On M-theory fourfold vacua with\\[.2cm] 
higher curvature terms} 

\vskip 1.5cm

\renewcommand{\thefootnote}{}

\begin{center}
 \normalsize
 \bf{Thomas W.~Grimm, Tom G.~Pugh, and Matthias Wei\ss enbacher}\footnote{\texttt{grimm,\ pught,\ mweisse @mpp.mpg.de}} 
\end{center}
\vskip 0.5cm

 \emph{ Max Planck Institute for Physics, \\ 
        F\"ohringer Ring 6, 80805 Munich, Germany} \\[0.25cm]

\end{center}

 \vskip 1.5cm
\renewcommand{\thefootnote}{\arabic{footnote}}

\begin{center} {\bf ABSTRACT } \end{center}

We study solutions to the eleven-dimensional supergravity action, including terms quartic and cubic in 
the Riemann curvature, that admit an eight-dimensional compact space. The internal background
is found to be a conformally K\"ahler manifold with vanishing first Chern class. The metric solution, however,
is non-Ricci-flat even when allowing for a conformal rescaling including the warp factor. This deviation is due to the possible non-harmonicity
of the third Chern-form in the leading order Ricci-flat metric. We present a systematic derivation of the background
solution by solving the Killing spinor conditions including higher curvature terms. These 
are translated  into first-order differential equations for a globally defined real two-form and complex four-form 
on the fourfold. We comment on the supersymmetry properties of the described solutions.

\end{titlepage}

\newpage

\setcounter{page}{1}
\setlength{\parskip}{9pt} 

\section{Introduction and summary}

The study of M-theory on eight-dimensional compact manifolds is of both conceptional 
as well as phenomenological interest. On the one hand, this compactifications allow the dynamics of three-dimensional effective theories with various amounts of 
supersymmetry to be investigated. 
On the other hand, the M-theory to F-theory limit can be used to 
lift the three-dimensional theories to four space-time dimensions for a certain class 
of eight-dimensional manifolds \cite{Vafa:1996xn}. From a phenomenological 
point of view, compactifications in which the 
effective theory preserves only small amounts of supersymmetry
are of particular interest. For example, compactifications of M-theory and F-theory preserving
four supercharges allow for background fluxes that can induce a  
four-dimensional chiral spectrum.

The aim of this note is to study vacua of eleven-dimensional supergravity on 
compact eight-dimensional manifolds $\cM_8$ including the known higher derivative 
terms to the action. More precisely, our starting point will include 
terms admitting eight derivatives and are fourth and third order in the eleven-dimensional
Riemann curvature $\hat \cR$, i.e.~schematically of the form $\hat \cR^4$ and $\hat \cR^3 \hat G^2$, where $\hat G$
is the field strength of the M-theory three-form. The terms fourth order in $\hat \cR$
are known since the works \cite{Duff:1995wd,Green:1997di,Green:1997as,Kiritsis:1997em,Russo:1997mk,Antoniadis:1997eg,Tseytlin:2000sf}, 
while recently the third order terms involving $\hat G$
have been analyzed in \cite{Liu:2013dna}. Given this action we introduce an Ansatz for the 
background metric and fluxes capturing corrections expanded in 
powers of $\eppr \propto \ell_{M}^{3}$, where $\ell_M$ is the 
eleven-dimensional Planck length. This Ansatz includes a warp-factor
as well as a shift of the internal metric at order $\eppr^2$~\cite{Becker:2001pm}.  
The field equations pose second order differential constraints on the shifted internal 
metric which we are able to solve explicitly. The internal manifold turns 
out to have still vanishing first Chern class, but the metric background 
has to be chosen to no longer be Ricci flat. At order $\eppr^2$ the deviation 
from Ricci-flatness is measured by the warp-factor and the non-harmonic part of the third Chern form 
$c_3^\tbzero$
on $\cM_8$ evaluated in the zeroth order, Ricci-flat metric.

In order to systematically find an explicit solution and 
analyze its supersymmetry properties we also study the eleven-dimensional
supersymmetry variations. Unfortunately, these are not known to 
the required order to give a complete check of the preservation of 
three-dimensional $\cN=2$ supersymmetry corresponding to four supercharges. 
It was, however, argued in \cite{Lu:2003ze,Lu:2004ng} that the eleven-dimensional 
gravitino variations have to include certain seven-derivative couplings 
involving three Riemann curvature tensors. Evaluated for the background Ansatz this
induces modified Killing spinor equations for a globally defined 
spinor on $\cM_8$ that has to exist in order to have a supersymmetric solution. 
We show that the integrability condition on these
Killing spinor equations yields the modified Einstein equations at 
order $\eppr^2$. Furthermore, we use the globally defined spinor
to introduce a globally defined real two-form $J$ and complex 
four-form $\Omega$. The Killing spinor equations translate into 
first order differential constraints on these forms, which imply 
that the metric is (conformally) K\"ahler. In fact, this formulation 
allows us to give a simple derivation of the $\eppr^2$ correction 
to the internal metric found by solving the Einstein equations. 
Our results can 
also be reformulated in terms of torsion classes on an $SU(4)$
structure manifold. We find that, upon separating the conformal rescaling of 
the internal metric, only the torsion 
form $\cW_5$ in $d \Omega = \overline{\cW}_5 \wedge \Omega$
is non-vanishing but exact. At the two-derivative level 
eleven-dimensional supergravity on $SU(4)$ structure manifolds
has recently been studied in \cite{Prins:2013wza}.

It should be stressed that the first part of our analysis 
closely parallels the seminal papers \cite{Becker:1996gj, Becker:2001pm}. 
In particular, the derivation of the equations of motion 
satisfied by the background is in accordance with 
\cite{Becker:2001pm}. We are, in addition, able to explicitly 
solve these conditions and give a geometric interpretation 
of the result. The fact that the metric is no longer Ricci 
flat when higher derivative couplings and $\alpha'$-corrections 
are taken into account is a classical result for Calabi-Yau manifolds without background fluxes 
in string theory \cite{Nemeschansky:1986yx} and has been recently investigated for 
$Spin(7)$ and $G_2$ compactifications \cite{Becker:2014rea}. It is gratifying to observe that this result indeed 
carries over to warped Calabi-Yau fourfold compactifications 
with fluxes of eleven-dimensional supergravity. 
To fully check supersymmetry, however, it would be interesting to show that the 
proposed gravitino variation is complete. Furthermore, it is still an open problem 
to derive the three-dimensional effective action including fluctuations 
around the presented background. 
If the resulting three-dimensional action carries the properties of 
a $\mathcal{N} = 2$ supergravity theory, this would give a further test for the supersymmetry 
of this background. We hope to present the derivation of the effective action 
in a forthcoming publication \cite{toappear} extending the results of \cite{Grimm:2013gma,Grimm:2013bha,Junghans:2014zla}.

The paper is organized as follows. In section \ref{warped_background}
we present the Ansatz for the metric and the background fluxes and 
give the equations satisfied by the appearing functions. We then solve 
the internal Einstein equations finding corrections to the metric. 
The gravitino variations are analyzed in section \ref{SusyAnalysis}. We derive 
the modified Killing spinor equations and translate the conditions 
into first order differential equations for $J,\Omega$. We comment on 
the compatibility with the Einstein equations and the implications for supersymmetry.
Useful identities and a summary of our conventions are 
supplemented in appendix \ref{Conventions}.

\section{Warped background solutions to eleven-dimensional supergravity} \label{warped_background}

In the following we will determine a bosonic solution to 
eleven-dimensional Einstein equations in the presence of 
higher curvature corrections and background fluxes. We will 
explicitly solve the Einstein equations finding a correction
to the internal Calabi-Yau metric. Supersymmetry properties of
this solution will be discussed in section \ref{SusyAnalysis}.

\subsection{The eleven-dimensional action}

Recall that the bosonic spectrum of eleven-dimensional
$\cN=1$ supergravity consists only of the metric $\hat g_{MN}$ and 
a three-form $\wh C$. We denote the field strength 
of $\hat C$ by $\wh G = d \wh C$ and note that the hats on the 
symbols indicate that we are dealing with eleven-dimensional fields, with indices raised and lowered with $\wh g_{MN}$. 

The dynamics of the fields is determined by  
the bosonic part of the $\cN=1$ supergravity action given by
\beq \label{S11expansion}
  S^{(11)} = S_{\rm class} + \eppr^2 S_{\wh R^4} +  \eppr^2 S_{\wh R^3 \wh G^2} + \ldots \ . 
\eeq
Here we have introduced the expansion parameter $\eppr$ given by
\beq
\eppr^2 = \fr{( 4 \pi \k_{11}^2)^\fr23}{(2 \pi)^4 3^2 2^{13}}  \ , 
\eeq
which is proportional to sixth power of the eleven-dimensional Planck length.
For the following analysis the relevant terms in \eqref{S11expansion} are, 
firstly, the classical two-derivative action \cite{Cremmer:1978km}
\beq
S_{\rm class} = \fr{1}{2 \k_{11}^2} \int
\wh R \wh * 1 - \fr12 \wh G \we \wh * \wh G - \fr16 \wh C \we \wh G \we \wh G \ ,
\eeq
where $\wh R$ is the Ricci scalar. Secondly, $S_{\wh R^4}$ 
denotes the terms quartic in the Riemann 
curvature and given by  \cite{Duff:1995wd,Green:1997di,Green:1997as,Kiritsis:1997em,Russo:1997mk,Antoniadis:1997eg,Tseytlin:2000sf}
\beq \label{SR4}
  S_{\wh R^4} = \fr{1}{2 \k_{11}^2} \int  (\wh t_8 \wh t_8 - \fr1{24} \wh \e_{11} \wh \e_{11} ) \wh R^4 \wh * 1 +  3^2 2^{13} \wh C \we \wh X_8\ .
\eeq
The explicit form of the various terms in \eqref{SR4} is given in appendix \ref{Conventions}. 
It is believed that these are all terms quartic in the Riemann tensor at this order in $ \eppr $.
The terms at higher order in $\wh G$ and $\eppr$, such as $S_{\wh R^3 \wh G^2}$,  
will not be needed in what follows as their 
contribution is higher order in $\eppr$ when evaluated on the ansatz we will make.

\subsection{Ansatz for the vacuum solution} \label{BackAnsatz}

We now consider solutions for which the internal space is a compact eight-dimensional manifold $\cM_8$
and the external space is $\bR^{2,1}$. 
At lowest order in $\eppr$ the solution takes the form 
\ba
 d \wh s^2 & = \wh g_{MN} dx^M dx^N = \h_{\m\n} dx^\m dx^\n + g^\tbzero_{m n} dy^m dy^n + \cO ( \eppr ) \ , & 
 \wh G &=0+\cO ( \eppr )\ ,
\ea
where $\m = 0,\ldots,2$ and  $m = 1,\ldots,8$. The Einstein equations imply Ricci-flatness of 
the internal space $ R^\tbzero_{m n} =0 $. In fact, 
together with the supersymmetry conditions requiring the preservation of four supercharges,
one infers that the internal manifold is Calabi-Yau 
and thus admits a nowhere vanishing K\"ahler form $J_{mn}^\tbzero$ and 
a holomorphic (4,0)-form $\O_{mnrs}^\tbzero$ that are harmonic. 

Having deduced this lowest order solution we can then work to second 
order in $\eppr$ by considering the field equations of the $\eppr$ corrected action. 
To solve the corrected Einstein equations we make an Ansatz for the metric \footnote{Note that an alternative ansatz with AdS external space can also be analysed. However, this is not compatible with the lowest order supersymmetry conditions on the flux combined with the second order equations of motion.}
\ba \label{MetricAnsatzeps2}
 d \wh s^2 &= e^{\eppr^2  \F^\tbtwo } ( e^{-2 \eppr^2 W^\tbtwo}   \h_{\m\n} dx^\m dx^\n  +  e^{  \eppr^2 W^\tbtwo}  g_{mn} dy^m dy^n )\ + \cO(\eppr^3) ,
\ea
where 
\ba 
  g_{mn} = g^\tbzero_{m n} + \eppr^2 g^\tbtwo_{m n} + \cO(\eppr^3)\, .
  \label{gexp}
\ea
Here $  \F^\tbtwo $, $W^\tbtwo$, $g^\tbzero_{m n}$ and $g^\tbtwo_{m n}$ depend only on the internal coordinates $y^m$ in the background. 
The function $ \F^\tbtwo$ is an overall Weyl rescaling that we will discuss in more 
detail below, while $W^\tbtwo$ is known as the warp-factor.
At this order in $\eppr$ a background four-form field strength must also be included. Following \cite{Becker:2001pm} we make the Ansatz 
\ba \label{FluxAnsatz}
 \wh G_{mnrs} & = \eppr G_{mnrs}^\tbone + \cO(\eppr^3) \, , &
  \wh G_{\m\n\r m} &= \e_{\m \n \r} \pa_m e^{ -3 \eppr^2 W^\tbtwo} + \cO(\eppr^3)  \,  ,
\ea
where $G^\tbone$ is a background four-form flux on the internal manifold $\cM_8$
that is harmonic with respect to $g^\tbzero_{mn}$.
Let us note that the term linear in $\eppr$ appearing in
$\wh G_{mnrs}$ has the correct mass dimensions such that 
the background flux $G_{mnrs}^\tbone$ integrates to a dimensionless number. In 
fact $T_{\rm M2} \int_{\cC_4} \wh G$ has to be dimensionless and 
the inverse M2-brane tension $T_{\rm M2}^{-1}$ is proportional to $\eppr$. We do not include a $\eppr^2$ term in the 
Ansatz for $\wh G_{mnrs}$, since it can be shown to either decouple or to give contributions at only $\cO(\eppr^3)$ in 
the following evaluations. 

\subsection{Equations determining the solution}

The functions appearing in our ansatz may then be constrained by substituting into the 
eleven-dimensional equations of motion. The solution is found by expanding 
each of the equations of motion in powers of $\eppr$ and inferring the 
respective constraints \cite{Becker:2001pm}. 

To begin with, we note that the equations of motion of $\wh C$ and the eleven-dimensional
Einstein equations derived from \eqref{S11expansion} do not decouple at first. However, combining the $\wh C$ equation
with the external Einstein equations one infers that $G^\tbone$ in the Ansatz \eqref{FluxAnsatz} 
is self-dual in the Calabi-Yau background, i.e.
\beq \label{selfdual}
     \eppr G^\tbone  = \eppr *^\tbzero G^{\tbone} + \cO(\eppr^3)\, ,
\eeq
where one uses that $\cM_8$ is compact. By using \eqref{selfdual} the second order equation of motion of $\wh C$  implies the warp-factor equation 
\ba \label{warpfactoreq}
  & \Delta e^{ 3 \eppr^2 W^\tbtwo}  + \tfr{1}{4! 2}  \eppr^2  G^\tbone_{m n r s}  G^{\tbone m n r s}
- \tfr{3^2 2^{13}}{8!}\eppr^2 \e^{m_1 \ldots m_8} X_{8\, m_1 \ldots m_8} + \cO(\eppr^3) = 0\ , 
\ea
where the Laplacian $\Delta =  \na_m  \na^m$, the $X_8$, and the contractions of $G^\tbone_{m n r s} $ are
evaluated using $g_{m n}$ given in \eqref{gexp}.
We stress that with the above Ansatz \eqref{FluxAnsatz} the corrections to the $\wh C$ equation of motion 
\eqref{selfdual} and \eqref{warpfactoreq} 
from $S_{\wh R^3 \wh G^2} $ in \eqref{S11expansion} give contributions at least of 
order $\eppr^3$. At this order not all  
higher curvature contributions are known. 
Therefore, these conditions give constraints only to
order $\eppr^2$. This indicates consistency of our Ansatz for the 
warp-factor and implies that lower $\eppr$ powers in the solution 
to \eqref{warpfactoreq} would be constants. 
Moreover, at this order in $\eppr$ the metric used in \eqref{warpfactoreq} 
is only $g^\tbzero_{m n}$.
Integrating \eqref{warpfactoreq} over the internal manifold $\cM_8$ one infers that, 
in the absence of localized sources, a non-trivial background flux $\til G^\tbone_{m n r s}$ is 
required by consistency for a manifold with $\int_{\cM_8} X^{\tbzero}_8 \neq 0$.

Next we use the Ansatz \eqref{MetricAnsatzeps2} and \eqref{FluxAnsatz}, along with the 
the constraints \eqref{selfdual} and \eqref{warpfactoreq},
to rewrite the Einstein equations into a simple form. 
Firstly, we expand 
\ba
   R_{mn} \equiv R( g^\tbzero_{rs} + \eppr^2 g^\tbtwo_{rs} )_{mn} = R^\tbzero_{mn} + \eppr^2 R^{\tbtwo}_{mn} 
\ea
which defines $R^{\tbtwo}_{mn}$.
Using this abbreviation the internal part of the eleven-dimensional 
Einstein equations can be rewritten as
\ba \
R^\tbtwo_{mn} - \tfr12 g^\tbzero_{m n}g^{\tbzero \, r s }  R^\tbtwo_{r s } +  768 J^\tbzero_{m}{}^r J^\tbzero_{n}{}^s \na_r \na_s Z  - \tfr92  \na_m \na_n \F^\tbtwo   + \tfr92 g^\tbzero_{m n} g^{\tbzero \, r s } \na_r \na_s \F^\tbtwo =0 \, ,
\label{intEinstein}
\ea
where $J^\tbzero_{m}{}^n =J^\tbzero_{mp} g^{\tbzero pn}$ is the complex structure on the underlying Calabi-Yau manifold. 
The conditions \eqref{selfdual} and \eqref{warpfactoreq} are used to cancel all 
flux dependence in \eqref{intEinstein} and ensure that the Einstein equations involving $\wh R_{m\mu}$ 
are automatically satisfied at the order considered.
The external part of the Einstein equations takes the form
\ba \label{exEinstein}
R^\tbtwo_{mn} g^{\tbzero \, m n} - 9 g^{\tbzero \, m n  } \na_m \na_n \F^\tbtwo =0 \, .
\ea
The derivation of \eqref{intEinstein} and \eqref{exEinstein} is rather lengthy and requires
the use of the identities summarized in appendix \ref{Conventions}. Furthermore, we 
have used Ricci-flatness $R^{\tbzero}_{mn}=0$ for the lowest order 
part of the Riemann tensor to simplify the result. 
In these expressions the scalar $Z$ is proportional to the six-dimensional 
Euler density and is given by 
\ba \label{Zdef}
Z &= *^\tbzero ( J^\tbzero \we c_3^\tbzero )  = \tfr1{12} (R^\tbzero_{m n}{}^{ r s}  R^\tbzero_{r s}{}^{ t u}R^\tbzero_{tu}{}^{ m n} - 2 R^\tbzero_m{}^r{}_n{}^s R^\tbzero_r{}^t{}_s{}^u R^\tbzero_t{}^m{}_u{}^n) \, ,
\ea
where $c_3^\tbzero$ the third Chern form evaluated in the metric $g_{mn}^\tbzero$
given explicitly in \eqref{c3-explicit}.
%
Tracing the internal part of the Einstein equation and demanding compatibility with the external part then fixes 
\ba
 \F^\tbtwo &= - \tfr{512}{3} Z\, , & 
 R_{mn}^\tbtwo &=  - 768 (J^\tbzero_{m}{}^r J^\tbzero_{n}{}^s \na_r \na_s Z +  \na_m \na_n Z )\ .
\label{EinsteinEqn}
\ea
In other words, the solution indeed requires the presence of a non-trivial 
eleven-dimensional Weyl rescaling involving the higher curvature terms.

\subsection{Solving the modified Einstein equation} \label{SolvingEinstein}

In order to solve \eqref{EinsteinEqn} we follow a technique equivalent to that 
shown in \cite{Nemeschansky:1986yx}. We begin by noting that as $c^\tbzero_3$ is real and closed 
but not co-closed with respect to the K\"ahler metric $g_{m n}^\tbzero$. This 
means that it may be expanded as 
\ba \label{def-F}
c_3^\tbzero = H c_3^\tbzero + i \pa^\tbzero \bar \pa^\tbzero F
\ea 
where $H$ indicates the projection to the harmonic part with respect to the metric $g_{mn}^\tbzero$. This equation 
defines a co-closed $(2,2)$-form $F$ that 
will be key to the following discussions.\footnote{The 
harmonicity of Chern forms has been also discussed in the mathematical literature 
and lead to the introduction of the Bando-Futaki character \cite{Bando}, which is however 
trivially vanishing in the Calabi-Yau case.} Then by using \eqref{Zdef} we see that 
\ba
\label{Zexapanded}
Z = *^\tbzero ( J^\tbzero \we H c_3^\tbzero) + \tfr{1}{4} \D^\tbzero *^\tbzero ( J^\tbzero \we J^\tbzero \we F) 
\ea
where $*^\tbzero ( J^\tbzero \we H c_3^\tbzero)$ is constant over the internal 
space as a result of the harmonic projection. We are now in the position to use these quantities to solve
\eqref{EinsteinEqn}  for a metric correction at order $\eppr^2$.
The explicit solution is given by 
\ba
\label{g2Expression}
g_{mn}^{\tbtwo} = 384 (J^\tbzero_{m}{}^r J^\tbzero_{n}{}^s \na_r^\tbzero \na_s^\tbzero  +  \na_m^\tbzero \na_n^\tbzero  ) *^\tbzero ( J^\tbzero \we J^\tbzero \we F) \ ,
\ea
where $F$ is the four-form introduced in \eqref{def-F}.
Clearly, one can now explicitly check that \eqref{g2Expression} solves \eqref{EinsteinEqn}.\footnote{Recently, it was pointed out in \cite{Junghans:2014zla}
that a redefinition of the metric background $g_{m n} =  g^{\tbzero}_{m n} - 768 \eppr^2 J^\tbzero_{m}{}^r ( \ast^\tbzero c_3^{\tbzero})_{r n} $ 
trivializes the kinetic terms for the vectors obtained from $\hat G$ in the three-dimensional 
effective action. This interesting observation, however, has to be contrasted with 
the fact that this shift is not a 
solution to the Einstein equations at order $\eppr^2$. } In the 
next section we will show by introducing globally defined forms on $\cM_8$ how one is naturally 
lead to the solution \eqref{g2Expression}.

\section{Killing spinor equations and globally defined forms} \label{SusyAnalysis}

In this section we comment on the supersymmetry properties of the solution 
introduced in section \ref{warped_background}. This is a challenging task, since 
the supersymmetry variations are not fully known at the desired order $\eppr^2$.
Following a strategy used in \cite{Lu:2003ze,Lu:2004ng} we will be able to extract at least partial 
information about the supersymmetry properties by studying the 
Killing spinor equations at order $\eppr^2$. Furthermore, we will then 
translate these equations into differential conditions on the globally defined 
forms $J$ and $\Omega$ on $\cM_8$. This will lead to a stepwise derivation of 
the correction \eqref{g2Expression}.

To set the stage of our study, let us note that we assert that at quadratic order in $\eppr$ the eleven-dimensional gravitino variation 
is given by 
\ba
\label{GravitinoVariation}
\d \wh \p_M &= \wh \na_M  \wh \e  - \tfr{1}{288} \wh G_{NRST} \wh \G_{M }{}^{NRST} \wh \e + \tfr{1}{36} \wh G_{M N R S} \wh \G^{NRS} \wh \e \nn \\
& \quad +  \tfr{128}{3} \eppr^2 \wh \na_N \wh Z \wh \G_M{}^N  \wh \e
  - 48 \eppr^2 \wh \na^N \wh R_{M R N_1 N_2} \wh R_{N S N_3 N_4} \wh R^{RS}{}_{N_5 N_6} \wh  \G^{N_1 \ldots N_6}  \wh \e +  \cO( \eppr^2)\ ,
\ea
where the remaining order $\eppr^2$ terms vanish on the backgrounds we consider. 
Here $\wh Z$ is proportional to the six-dimensional Euler 
density in eleven dimensions and is given by 
\ba
 \wh Z = \tfr1{12} (\wh R_{M N}{}^{RS}  \wh R_{RS}{}^{ TU} \wh R_{TU}{}^{MN} - 2 \wh R_M{}^R{}_N{}^S \wh R_R{}^T{}_S{}^U \wh R_T{}^M{}_U{}^N)\ .
\ea
This form of the gravitino variation is compatible with the terms that are necessary in \cite{Lu:2003ze,Lu:2004ng}.
In other words, we will see below that the Killing spinor equations derived 
from \eqref{GravitinoVariation} are compatible with the Einstein equations up to order $\eppr^2$.
Remarkably, the terms in \eqref{GravitinoVariation} also appear in the gravitino 
variations deduced by eleven-dimensional Noether coupling in \cite{Hyakutake:2006aq}.

\subsection{Dimensional reduction of the supergravity variations} \label{DimRedSUGRA}

We next dimensionally reduce the supersymmetry variations \eqref{GravitinoVariation}
on the background introduced in section \ref{warped_background}.
To begin with, we decompose the eleven-dimensional supersymmetry parameter 
and gamma matrices in a way that is compatible with our Ansatz as 
\ba
  \wh \e & = e^{-\fr12  \eppr^2 W^\tbtwo} \e \otimes \eta \ , &
 \wh \G_\m  & = e^{ \frac12 \eppr^2 \F^\tbtwo - \eppr^2 W^\tbtwo  }  \g_\m \otimes \g^9 \ , &
  \wh \G_m & =  e^{ \frac12 \eppr^2 \F^\tbtwo + \frac12 \eppr^2 W^\tbtwo  }  \id \otimes \g_m \ ,
\ea
where $\e$ is a spinor in the three-dimensional external space and $\eta$ 
is a no-where vanishing spinor on $\cM_8$.
The spinor $\eta$ is chosen to satisfy  $\g^9 \h  = \h$,  $\h^{\dagger}\h  =1$ and  $\h^{ T}\h =0$.

Substituting this decomposition along with the reduction ansatz \eqref{MetricAnsatzeps2} and 
\eqref{FluxAnsatz} into \eqref{GravitinoVariation} we find for the internal gravitino variation 
\ba
\d \wh \p_m =&\ e^{ - \fr12 \eppr^2 W^\tbtwo} \e \otimes \nabla_m  \h   - \tfrac{1}{288} \eppr G^{\tbone}_{n r s t } \e \otimes \g_m{}^{n r s t}  \h  + \tfrac{1}{36} \eppr  G^{\tbone}_{mnpq} \e \otimes \gamma^{npq}  \h \nn \\ &
  - 48 \eppr^2 \na^n R_{m r m_1 m_2} R_{n s m_3 m_4} R^{r s}{}_{m_5 m_6} \e \otimes \g^{m_1 \ldots m_6} \h  \nn \\
& + \tfr{128}{3} \eppr^2\na_n Z \e \otimes \g_m{}^n \h + \tfrac14 \eppr^2 \na_n \F^\tbtwo \e \otimes \g_m{}^n \h  + \cO(\eppr^3) = 0\ ,
\ea
and for the external gravitino variation 
\ba
\d \wh \p_\m  =&\  e^{ - \fr12 \eppr^2 W^\tbtwo}  \na_\m \e \otimes  \h  - \eppr \tfrac{1}{288} G^{\tbone}_{mnpq} \g_\m \e \otimes \gamma^{mnpq}  \h \nn \\
& -  \tfr{128}{3} \eppr^2 \na_n Z \g_\m \e  \otimes \g^n \h - \tfrac14 \eppr^2 \na_n \F^\tbtwo \g_\m \e \otimes \g^n \h   + \cO(\eppr^3) = 0\ .
\ea
These equations can then be satisfied if at lowest order in $\eppr$ if the background is Calabi-Yau, as already noted at the 
beginning of section \ref{BackAnsatz}, and one has $\na_\m \e = 0$. At linear order in $\eppr$ one finds the condition 
\ba
\label{HigherSusyConds1}
G^\tbone_{m n r s} \g^{n r s} \eta &= 0
\ea
Finally, at second order in $\eppr$ one finds that \eqref{EinsteinEqn} has to be satisfied and 
$\eta$ obeys the Killing spinor equation
\ba \label{HigherSusyConds2}
\na_m \h &=  - 384 \eppr^2 J^\tbzero{}_{m}{}^n \na_n Z_{r s} \g^{r s} \h + \cO(\eppr^3) \ ,&
Z_{rs} &= \tfr12 (* c_3^\tbzero)_{r s} 
\ea
where  $ J^{\tbzero r s} Z_{r s} = Z $.

\subsection{Differential conditions on the globally defined forms} \label{Killingequation}

Using the spinor $\h$ one can introduce a globally defined no-where vanishing real two-form $J$
and a complex four-form $\Omega$. This is a familiar strategy for manifolds with reduced 
structure group. The case of having $SU(4)$ structure was discussed in \cite{Prins:2013wza,Prins:2013koa}.
Concretely, we use $\h$ to construct the forms
\ba \label{def-JOmega}
J_{m n}  =  i \h^\dagger \g_{m n} \h\ , \qquad 
\O_{mnrs}  =  \h^T \g_{m n r s} \h\ . 
\ea
By using Fierz identities we see that these forms satisfy
\ba
\label{NormalisationConds}
J \we \O &= 0 \ ,&
J \we J \we J \we J &= \tfrac32 \O \we \bar \O \ .
\ea
The K\"ahler form $J_{m n}^\tbzero$ corresponding to the Ricci flat metric $g_{m n}^\tbzero$ is then the lowest order part of $J_{m n }$. 

We can now rewrite the supersymmetry conditions \eqref{HigherSusyConds1} and \eqref{HigherSusyConds2} 
using $J$ and $\Omega$. The constraint on the flux \eqref{HigherSusyConds1} implies that
\ba
 G^\tbone \we J^\tbzero   &= 0 \ , &
  & G^\tbone\ \text{is of type (2,2) in}\ J^{\tbzero\, n}_m\,   
\ea
where $J^{\tbzero\, n}_m$ is the complex structure of the underlying 
Calabi-Yau fourfold.
Furthermore, the Killing spinor equation \eqref{HigherSusyConds2} satisfied by $\h$ translates to 
the differential conditions
\ba
\label{CovDofJandOm}
\na_m J_{n r}  &= 0 + \cO(\eppr^3) \ , &
\na_m \O_{n r s t}  &= 6144 \eppr^2 J^\tbzero_m{}^p  \na_p^\tbzero   Z_{ [n}{}^q \O^\tbzero_{r s t ] q} + \cO(\eppr^3) 
\ea
Antisymmetrising in the indices then gives
\ba
d J  &= 0+ \cO(\eppr^3)  \ , &
d \O &= - 768 \eppr^2  d Z \we \O^\tbzero + \cO(\eppr^3) \ .
\label{dJdOmConds}
\ea
We can thus infer that the metric $g_{m n}$ including $\eppr^2$ corrections 
is still K\"ahler. In fact, the higher curvature terms only amount to introducing 
the non-closedness of $\O$ with a result proportional to $\O$ itself.
In fact, translated into torsion forms for an $SU(4)$-structure manifold (see, for example, \cite{Prins:2013wza,Prins:2013koa}), the only 
non-trivial torsion form is $\overline{ \cW}_{5} =- 768 \eppr^2  \bar \partial^\tbzero Z$, which is  exact.

Let us stress that the derivation of the Killing spinor equation makes use of the full internal space metric $\wh g_{MN}$. However, the overall Weyl rescaling and warp-factor terms precisely cancel and the resulting equation  \eqref{HigherSusyConds2} depends only on the metric $g_{mn}$ appearing in \eqref{MetricAnsatzeps2}. The $J$ and $\O$ appearing\eqref{dJdOmConds} are thus related to the metric $g_{mn}$. Clearly one could introduce a alternative $\til J$ and $\til \O$ related to rescaled metric $\wh g_{mn}$. This would induce new terms proportional to $\til J$ in $d \til J$ and $\til \Omega$ in $d \til \Omega$ will then 
be induced, since the gamma-matrices in \eqref{def-JOmega} are rescaled.

We can now use the condition that $g_{mn}$ is a K\"ahler metric and study the 
integrability condition of  \eqref{HigherSusyConds2}. 
Here the commutator $[\nabla_m,\nabla_n] \eta = \frac14 R_{mnrs} \gamma^{rs} \eta$ 
can be compared with the result obtained form \eqref{HigherSusyConds2}. This simply 
results in the condition
\ba
\fr14 R_{m n r s} \g^{rs} \h - 768 \eppr^2 J^\tbzero{}_{[m}{}^r \na^\tbzero_{n]} \na^\tbzero_r Z_{p q} \g^{pq} \h + \cO(\eppr^3) = 0  \ .
\ea
Contracting with $\eta^\dagger$ we see that this implies  
\ba \label{intcondition}
\tfr14 R_{m n r s} J^{r s} -  768 \eppr^2 J^\tbzero{}_{[m}{}^r \na^\tbzero_{n]} \na^\tbzero_r Z  + \cO(\eppr^3) = 0  \ .
\ea
As we know that $R_{m n r s} J^{r s} =  2 R_{m r n s} J^{r s}$ by the first Bianchi 
identity and that for a K\"ahler manifold $ J_{m}{}^{ p} R_{p n r s} = J_{n}{}^{ p} R_{m p r s} $ 
we then see that \eqref{intcondition} implies $R^{\tbzero}_{m n}=0$ at zeroth $\eppr$ order and 
the Einstein equations \eqref{EinsteinEqn} at order $\eppr^2$.

\subsection{Solving the equations for $J$ and $\O$}

We now wish to solve the equations \eqref{dJdOmConds} subject to the algebraic constraints \eqref{NormalisationConds}. To do this we begin by expanding these equations in $\eppr$ to find 
\ba
d J^\tbtwo & = 0 \, , & 
d \O^\tbtwo & = - 768 d Z \we \O^\tbzero \, . 
\ea
We may solve the constraint on $\O^\tbtwo$ by letting 
\ba
\O^\tbtwo &= \f \O^\tbzero + \r \, , &
&\text{where}&
d  \f &= - 768  d Z \, , &
d \r &=0 \, .
\label{OmegaExpansion}
\ea 
The (4,0) part of $\r$ can be absorbed into $\f \O^\tbzero$ so we may assume that $\r \we \bar \O^\tbzero = 0$. Similarly as $J^\tbtwo$ is a real d-closed 2-form on a Kahler manifold
\ba
J^\tbtwo &=  \s + i \pa^\tbzero \bar \pa^\tbzero \p \, ,&
&\text{where}&
d \s &= d^{\tbzero \dagger} \s = 0 \, . 
\label{JExpansion}
\ea 
Then considering the expansion of \eqref{NormalisationConds} we see that 
\ba
\label{expandedNormCond}
4 J^\tbtwo \we J^\tbzero \we J^\tbzero  \we J^\tbzero &= \fr32 (\O^\tbtwo \we \bar \O^\tbzero+\O^\tbzero \we \bar \O^\tbtwo)  \, , 
\ea 
and substituting \eqref{OmegaExpansion} and \eqref{JExpansion} into \eqref{expandedNormCond} we find
\ba
 \fr13 *(\s \we J^\tbzero  \we J^\tbzero \we J^\tbzero ) -   \D^\tbzero \p = 2 (\f + \bar \f)  \, ,
\ea
which implies that $d \D^\tbzero \p = 3072 d Z$. Considering this along with \eqref{OmegaExpansion} and using the expansion of $Z$ given by \eqref{Zexapanded} we see that we are lead to a solution for $J^\tbtwo$ and $\O^\tbtwo$ where 
\ba
J^\tbtwo & =  i 786 \, \pa^\tbzero \bar \pa^\tbzero *^\tbzero ( F \we J^\tbzero \we J^\tbzero) \, ,&
\O^\tbtwo & = - 192 \, \D^\tbzero *^\tbzero ( F \we J^\tbzero \we J^\tbzero) \O^\tbzero \, . 
\label{JOmTwoSols}
\ea
This shows that the internal space K\"ahler potential is shifted by a term proportional to F. The remaining forms $\r$ and $\s$ correspond to moduli which will be studied in \cite{toappear}.
Expanding the relationship 
\ba
g_{m n} = \tfr{i}{48} \O_{(m| r p t} \bar \O_{|n) s q u} J^{r s} J^{p q} J^{t u} \, , 
\ea
which may be demonstrated using the results of Appendix \ref{Conventions}, we find 
\ba
g_{m n}^\tbtwo = - J_{(m}^\tbzero{}^r J_{n) r}^\tbtwo + \fr12 J^{\tbzero r s } J_{rs}^\tbtwo g^\tbzero_{m n} - \fr{1}{48} \O_{(m| r s t}^\tbtwo \bar \O_{|n)}^\tbzero{}^{ r s t} - \fr{1}{48} \bar \O_{(m| r s t}^\tbtwo \O_{|n)}^\tbzero{}^{ r s t} \, , 
\ea
and using this we see that the correction to $J$ and $\O$  implies the metric correction \eqref{g2Expression} that solves \eqref{EinsteinEqn}.

The analysis presented here shows that the first order equations \eqref{dJdOmConds} on $J$ and $\O$, which are derived from the Killing spinor equations \eqref{HigherSusyConds2} are economically solved by \eqref{JOmTwoSols}. This then provides a solution to the second order equations  \eqref{EinsteinEqn} arising from the internal space Einstein equations. While we have no complete proof of the supersymmetry of this solution this result provides a necessary condition. Furthermore, as we expect that the lowest order supersymmetry carries over to the higher order analysis and we have made a general analysis of the corrections to the eleven-dimensional field equations, it seems natural to expect that further corrections to the gravitino variation \eqref{GravitinoVariation} vanish in the background presented. It would be interesting to continue to develop the Noether coupling analysis of \cite{Hyakutake:2006aq} to find the complete expression for the gravitino variation at order $\eppr^2$.

\vspace*{.5cm}
\noindent
\subsection*{Acknowledgments}

We like to thank Ralph Blumenhagen, Federico Bonetti, Akito Futaki, Daniel Junghans and Raffaele Savelli for
useful discussions and comments. This work was supported by a grant of the Max Planck Society. 

\begin{appendix}
\vspace{2cm} 
\noindent {\bf \LARGE Appendix}

\section{Conventions, definitions and identities} \label{Conventions}

We denote the total eleven-dimensional space indices by capital Latin letters $M,N,R,S,\dots$,the external  ones by  $\mu,\nu = 0,1,2$ and real indices of the internal space by $m,n,r,s=1,\dots, 8$. Quantities for which the indices are raised and lower with the total space metric carry a hat e.g.  $\wh G,\wh R$. Furthermore, the convention for the totally 
anti-symmetric tensor in Lorentzian space in an orthonormal frame is $\epsilon_{012...10} = \epsilon_{012}=+1$. The epsilon tensor in d dimensions then satisfies
\ba
\epsilon^{R_1\cdots R_p N_{1 }\ldots N_{d-p}}\epsilon_{R_1 \ldots R_p M_{1} \ldots M_{d-p}} &= (-1)^s (d-p)! p! 
\delta^{N_{1}}{}_{[M_{1}} \ldots \delta^{N_{d-p}}{}_{M_{d-p}]} \,, 
\ea
where  $s=0$ if the metric has Riemannian signature and $s=1$ for a Lorentzian metric.

We adopt the following conventions for the Riemann tensor of the internal space 
\ba
\G^r{}_{m n} & = \fr12 g^{rs} ( \pa_{m} g_{n s} + \pa_n g_{m s} - \pa_s g_{m n}  ) \, , &
R_{m n} & = R^r{}_{m r n} \, , \nn \\
R^{m}{}_{n r s} &= \pa_r \G^m{}_{s n}  - \pa_{s} \G^m{}_{r n} + \G^m{}_{r  t} \G^t{}_{s n} - \G^m{}_{s t} \G^t{}_{r n} \,, &
R & = R_{m n} g^{m n} \, , 
\ea
with equivalent definitions for the Riemann tensor on the total and external spaces. Perturbing the internal metric by $  g_{mn} = g^\tbzero_{m n} + \eppr^2 g^\tbtwo_{m n} $ the correction to the internal  Ricci tensor at $\mathcal{O}(\eppr^2)$ is then given by
\ba
R^\tbtwo _{m n} &=
\eppr^2 \nabla^\tbzero_r \nabla^\tbzero_{(m} g_{n)}^{\tbtwo r}  
 - \eppr^2 \frac{1}{2}   \nabla^{\tbzero r} \nabla^\tbzero_r g^\tbtwo_{mn} 
 - \eppr^2 \frac{1}{2}   \nabla^\tbzero_m \nabla^\tbzero_n g_{r}^{\tbtwo r}  \; .
\ea

The  scalar functions $ \wh t_8 \wh t_8 \wh R^4 $ and $ \wh \e_{11} \wh \e_{11} \wh R^4 $ are given by
\ba
\wh \e_{11} \wh \e_{11} \wh R^4 & =  \epsilon_{R_1 R_2 R_3 N_1 \ldots N_{8}} \epsilon^{R_1 R_2 R_3  M_1 \ldots M_{8} } \wh R^{N_1 N_2}{}_{M_1 M_2}  \wh R^{N_3 N_4}{}_{M_3 M_4}  \wh R^{N_5 N_6}{}_{M_5 M_6}  \wh R^{N_7 N_8}{}_{M_7 M_8}  \,,  \nn \\
 \wh t_8 \wh t_8 \wh R^4 &=  \wh t_8{}_{N_1 \ldots N_{8}} \wh t_8{}^{ R_3  M_1 \ldots M_{8} } \wh R^{N_1 N_2}{}_{M_1 M_2}  \wh R^{N_3 N_4}{}_{M_3 M_4}  \wh R^{N_5 N_6}{}_{M_5 M_6}  \wh R^{N_7 N_8}{}_{M_7 M_8} \, , 
\ea
with
\ba
\hat t_8^{N_1\dots N_8}   &= \fr{1}{16} \big( -  2 \left(   \wh g^{ N_1 N_3  }\wh g^{  N_2  N_4  }\wh g^{ N_5   N_7  }\wh g^{ N_6 N_8  } 
 + \wh g^{ N_1 N_5  }\wh g^{ N_2 N_6  }\wh g^{ N_3   N_7  }\wh g^{  N_4   N_8   }
 +  \wh g^{ N_1 N_7  }\wh g^{ N_2 N_8  }\wh g^{ N_3   N_5  }\wh g^{  N_4 N_6   }  \right) \nn \\
 & \quad +
 8 \left(  \wh g^{  N_2     N_3   }\wh g^{ N_4    N_5  }\wh g^{ N_6    N_7  }\wh g^{ N_8   N_1   } 
  +\wh g^{  N_2     N_5   }\wh g^{ N_6    N_3  }\wh g^{ N_4    N_7  }\wh g^{ N_8   N_1   } 
  +   \wh g^{  N_2     N_5   }\wh g^{ N_6    N_7  }\wh g^{ N_8    N_3  }\wh g^{ N_4  N_1   } 
\right) \nn \\
& \quad - (N_1 \lra N_2) -( N_3 \lra N_4) - (N_5 \lra N_6) - (N_7 \lra N_8) \big) \,. 
\ea
While the 8-form $X_8$ is given by 
\beq
X_8 =   \frac{1}{192}\left[\Tr (\wh{\mathcal{R}}^4) - \frac{1}{4}\left( \Tr (\wh{\mathcal{R}}^2) \right)^2\right]\,,
\eeq
where $ \wh{\mathcal{R}}^M{}_N  =  \frac{1}{2}\wh R^M{}_{N RS} dx^R\wedge dx^S $ and the 3rd chern form on the internal space may be expressed as 
\ba\label{c3-explicit}
 {c_3} =& - \fr1{48} R_{m_1 m_2 n_1 n_2}R_{m_3 m_4 n_3 n_4} R_{m_5 m_6 n_5 n_7} J^{n_2 n_3} J^{n_4 n_5} J^{n_6 n_1} dx^{m_1} \we \ldots \we d x^{m_6} \; .\nn
\ea

From the spinor bilinear $J$ we may form the projectors
\ba
 \Pi^{\pm \phantom{m} n }_{\phantom{\pm}m} & = \frac{1}{2} \left(\delta^{ \phantom{m} n }_{m} \mp i J^{ \phantom{m} n }_{m}  \right)\,, &
 & \text{where}&
 \Pi^{- \phantom{m} i }_{\phantom{+}m}  \Omega_{inrs} &=\Omega_{mnrs} \, , &
 \Pi^{+\phantom{m} i }_{\phantom{-}m}  \Omega_{inrs} &= 0 \,  ,
\ea
which satisfy 
\ba
\O_{m n r s} \bar \O^{t u v w} & = 4! \,  2^4 \Pi^{-}{}_{[m}{}^t \Pi^{-}{}_n{}^u \Pi^{-}{}_r{}^v \Pi^{-}{}_{s]}{}^w\, . 
\ea
as may be shown by using Fierz identities \cite{Tsimpis:2007sx}. Using these techniques we can also show that the remaining spinor bilinears on the internal space can be written as
\ba
\h^{\dagger} \gamma_{m n r s } \h &= -3 J_{[m n}J_{r s]}  \, , &
\h^{\dagger} \gamma_{m n r s t u} \h &= 15 i  J_{[m n}J_{rs}J_{tu]}    \, , &
\h^{\dagger} \gamma_{m n r s t u v w} \h &= 105 J_{[m n}J_{rs}J_{tu}J_{vw]}  \, , \nn 
\ea 
\vspace{-1cm}
\ba
   \eta^T \gamma_{p_1...p_d} \eta &= 0 \quad  \text{ where } d \neq 4 \, , &
   \eta^\dagger \gamma_{p_1...p_d} \eta &= 0 \quad  \text{where } d =\text{odd} \, . 
\ea

\end{appendix}



\end{document}